\documentstyle[epsfig]{aipproc}

\newcommand\psra{PSR~B1913+16}
\newcommand\psrb{PSR~B1534+12}

\begin{document}

\title{Relativistic Gravity \\ and Binary Radio Pulsars}

\author{Victoria M. Kaspi}
\address{Department of Physics and Center for Space Research\\
Massachusetts Institute of Technology\\Cambridge, MA 02139}

\maketitle

\begin{abstract}

Following a summary of the basic principles of pulsar timing, we
present a review of recent results from timing observations of
relativistic binary pulsars.  In particular, we summarize the status
of timing observations of the much celebrated original binary pulsar
\psra, draw attention to the recent confirmation of strong evidence
for geodetic precession in this system, review the recent measurement
of multiple post-Keplerian binary parameters for \psrb, and describe
the Parkes Multibeam survey, a major survey of the Galactic Plane
which promises to discover new relativistic binary pulsar systems.

\end{abstract}

\section*{Introduction} 

Not long after Einstein proposed his General Theory of Relativity, a
variety of experimental tests to be done with solar system objects was
suggested.  These included the measurement of the perihelion advances
of planets, the bending of light rays by the Sun, and radar echo
delays from planets.  However, such tests were limited by the fact
that the effects to be measured were tiny perturbations on a classical
description.  They only verified the theory in the ``weak-field''
limit, akin to studying a function by only considering its Taylor
expansion about zero.  The ``strong-field'' regime, in which GR
effects are more than a perturbation and a classical description is
grossly violated, probably at first appeared inaccessible to
Earth-bound observers.

The discovery of the first binary pulsar, \psra, by Hulse \& Taylor (1975)
\nocite{ht75} radically changed this situation.  This binary system,
consisting of two neutron stars in an eccentric 8~hr binary orbit, has
permitted precise tests of GR predictions for the first time in the
strong-field regime \cite{tw82,tw89}.  Thus far, GR has passed all 
tests with flying colours.

In this review, after an introduction to pulsars and pulsar timing, we
present the most recent results of observations of \psra, as well as
of \psrb, the second discovered binary pulsar system suitable for
sensitive GR studies.  We also describe a search for new pulsars that
is currently underway, and which promises to find more such objects.
For previous excellent reviews of relativistic binary pulsars and
their experimental constraints on strong-field relativistic gravity
see Taylor et al.  (1992) and Damour \& Taylor (1992).
\nocite{twdw92,dt92}

\section*{Radio Pulsars:  Some Background}

Pulsars are rotating, magnetized neutron stars.  They exhibit beams of
radio emission that can be observed, by a fortuitously located
astronomer, as pulsations, once per rotation period.  In the published
literature there are 708 pulsars known (but see section ``Parkes
Multibeam Survey'' below), all but a handful of which are in
the Milky Way, the remainder being in the Magellanic Clouds.  Known
pulse periods range from a few seconds down to 1.5~ms.  These pulse
periods are observed to increase steadily, indicative of spin-down due
to magnetic dipole radiation.  From the observed pulse period and rate
of spin-down of a pulsar, the magnitude of the dipole component of the
stellar magnetic field, as well as an age estimate, can be deduced.
See Lyne \& Smith (1998) \nocite{ls98} for a complete review of the
properties of radio pulsars.

For our purposes here, we need highlight only two properties of radio
pulsars: the stabilities of the radio pulse profile and the stellar
rotation.  By ``pulse profile'' we mean the result of the addition of
many (typically thousands) of individual pulses, by folding the
sampled radio telescope power output modulo the apparent pulse period.
Two examples of such pulse profiles are shown in Figure~1.  Average
profiles are observed to be stable in that the summation of any few
thousand consecutive pulses always results in the same pulse profile
for a given radio pulsar at a given observing frequency, even though
individual pulse morphologies vary greatly.  Currently there is no
theory to explain this observation; in radio pulsar timing it is
simply accepted as fact.  Less surprising perhaps is the observed
rotational stability.  In a reference frame not accelerating with
respect to the pulsar, the observed times of pulsations (or TOAs, for
times-of-arrival) are generally predictable with high precision, given
only the pulse period and spin-down rate.  This, we argue, is less
surprising than the profile stability because of the large stellar
moment of inertia and absence of external torques, in strong contrast
to accreting neutron stars whose rotation is much less stable [e.g.
Bildsten et al. 1997]. \nocite{bcc+97}

\begin{figure}[t]
\centerline{\epsfig{file=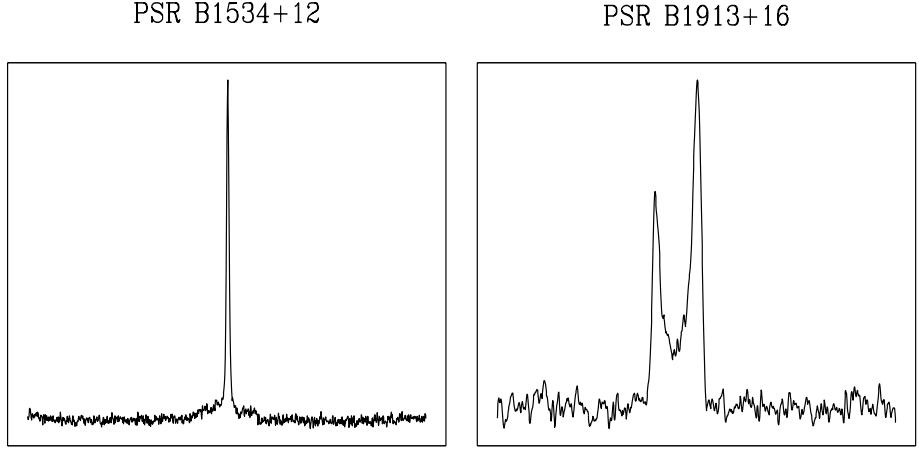,height=2.5in,width=5in}}
\caption{Average pulse profiles for \protect\psrb\ [after Kramer et al. 1998], \protect\nocite{kxl+98}
and \protect\psra\ (courtesy M. Kramer), both at radio frequencies near 1400~MHz.}
\label{fig:std}
\end{figure}

\section*{Pulsar Timing}

The combination of pulse profile and rotational stability makes a
radio pulsar useful as an extremely precise clock; in some cases the
stability of the pulsar-clock is comparable to those of the world's
best atomic time standards [e.g. Kaspi, Taylor \& Ryba
1994].\nocite{ktr94} However, the realization of this stability can
come only after effects extrinsic to the pulsar are accounted for.  In
particular, TOAs measured at an Earth-bound radio telescope must be
transformed to a reference frame that is not accelerating with respect
to the pulsar.  For this purpose, the solar system barycentre
reference frame is generally used.

Standard pulsar timing thus consists of observing a pulsar
at a radio telescope continuously over many cycles.  The start time of the 
observations is recorded with high precision, and the sampled telescope power
output is folded at the topocentric (i.e. apparent) pulse period.  The resulting
average pulse profile is cross-correlated
with a high signal-to-noise template (e.g. Fig.~\ref{fig:std}) 
in order to determine the
arrival time of the average pulse.  That time is then transformed to
the solar system barycentre.  This transformation can be summarized by the expression
\begin{equation}
t_{\rm SSB} =  t_{\rm O}  +  \Delta t_{\rm C} +  \Delta t_{\rm R}  +  \Delta t_{\rm E} 
+ \Delta t_{\rm S} + \Delta t_{\rm D},
\label{eq:t}
\end{equation}
where $t_{\rm SSB}$ is the pulse arrival time at the solar system barycentre
(typically in Baryncetric Dynamical Time),
$t_{\rm O}$ is the arrival time as observed at an Earth-bound radio telescope,
$\Delta t_{\rm C}$ is the difference between the observatory clock and a suitably
stable atomic time standard (such as Terrestrial Dynamical Time),
$\Delta t_{\rm R}$ is the Roemer delay, or the difference in arrival time of a pulse at the solar system barycentre
and at the observatory due to the geometric path length difference,
$\Delta t_{\rm E}$ is the Einstein delay due to (weak-field) GR effects in 
the solar system, and $\Delta t_{\rm S}$ is the so-called ``Shapiro delay,'' which
depends logarithmically on the impact parameter of the Earth-pulsar and Earth-Sun
line of sights.
Note that $\Delta t_{\rm R}$, $\Delta t_{\rm E}$ and $\Delta t_{\rm S}$ require 
precise knowledge of the sky coordinates of the pulsar;
this is turned around so that if observations of the source are
available over at least one year, the known motion of the
Earth in its orbit permits the measurement of the pulsar's 
coordinates with high precision.
The last term, $\Delta t_{\rm D}$, is an observing frequency-dependent 
term that accounts for
the dispersion of radio waves in the ionized interstellar medium 
according to the cold
plasma dispersion law.  The delay term is proportional to DM$/f^2$, 
where DM is the
dispersion measure, or integrated electron density along the line 
of sight, and $f$ is 
the observing frequency.  The measured DM, together with a model for
the distribution of free electrons in the Galaxy [e.g. Taylor \& Cordes
1993], provides an estimate of the distance to a pulsar.
Details of all the above terms can be found in various
references [e.g. Manchester \& Taylor 1977].\nocite{mt77}

The above procedure for timing a pulsar of interest is repeated
typically on a bi-weekly or monthly basis, so that the spin and
astrometric parameters are improved in an iterative fashion: the
squares of the residual differences between the initial
model-predicted TOAs and the observed TOAs are minimized by varying,
and hence improving, the model parameters.  The transformation and
subsequent determination of the five optimal spin and astrometric
parameters (the period $P$, its rate of change $\dot{P}$, two sky
coordinates and DM) are done using a publically available software
package, {\tt tempo}, which consists of several thousand lines of
Fortran code\footnote{\tt
http://pulsar.princeton.edu/tempo/index.html}.  Note that by using
TOAs, as opposed to measuring the pulse period at each observing
epoch, the timing analysis is coherent in the sense that every
rotation of the neutron star is accounted for.

\section*{Timing Binary Pulsars}
\label{sec:binary}

If the pulsar is in a binary system, its motion about the binary
centre of mass will cause regular delays and advances in observed TOAs
just as the Earth's motion around the Sun does.\footnote{ Although
most non-degenerate stars are in binary systems, most pulsars are
isolated because supernova explosions usually disrupt binaries.  See
Bhattacharya \& van den Heuvel (1991) \nocite{bv91} for a review of
the circumstances under which binary pulsars form.}  Classically, five
additional parameters are required to describe and predict pulse
arrival times for binary pulsars, in addition to the five spin and
astrometric parameters.  Conventionally the five Keplerian parameters
are the orbital period $P_b$, the projected semi-major axis $a \sin
i$, where $i$ is the inclination angle of the orbit, the orbital
eccentricity $e$, the longitude of periastron $\omega$ measured from
the line defined by the intersection of the plane of the orbit and the
plane of the sky, and an epoch of periastron $T_0$.  Only the
projected semi-major axis is measurable, as pulsar timing is only
sensitive to the radial component of the pulsar's motion.  Therefore,
the component masses cannot be uniquely determined.  Note that under
certain circumstances, even in a classical system, the five Keplerian
parameters may be insufficient to fully describe the orbit; for
example, in the binary pulsar PSR~J0045$-$7319, classical spin-orbit
coupling induces post-Keplerian dynamical effects, a result of the
quadrupole moment of the pulsar's rapidly rotating B-star companion
\cite{lbk95,kbm+96}.

In some binary systems, particularly double neutron star binaries,
relativistic effects must also be taken into account in order to model
the binary orbit and hence observed TOAs properly.  A list of the
known double neutron star binaries is given in Table~1.  The only
non-classical post-Keplerian (PK) effects to have been measured in a
binary pulsar system thus far are: the rate of periastron advance
$\dot{\omega}$, the combined effects of relativistic Doppler shift and
time dilation $\gamma$ (equivalent to the solar system Einstein delay
-- see Eq.~\ref{eq:t}), the rate of orbital decay $\dot{P_b}$, and $r$
and $s$, the two parameters describing the Shapiro Delay, or the
observed pulse time delay due to the bending of space-time near the
pulsar companion, important for highly inclined orbits (equivalent to
$\Delta t_{\rm S}$ in Eq.~\ref{eq:t}).  The relativistic
post-Keplerian parameters measured in each of the known double neutron
star binaries are given in Table~1.  The systems for which tests of
theories of relativistic gravity are possible are indicated by bold
type: these are binaries for which $N$ post-Keplerian parameters are
measurable, where $N>2$.  These systems permit $N-2$ tests of gravity,
as the first two parameters determine the masses of the two components.

Overall, the suitability of a binary pulsar system for tests of GR or
other theories of gravity is determined by a number of factors,
including orbital period, orbital eccentricity, orbital inclination
angle, the morphology of the pulse profile (narrower pulses permit
higher measurement precision) and of course, the pulsar's radio flux.
For example, PSR B2127+11C, though in a binary system that is superb
for testing GR \cite{pakw91}, is faint (it was discovered in a deep
search of the globular cluster M15) and has thus
far not permitted any tests of GR.

\begin{table}
%\begin{center}
\begin{tabular}{ccccc}
\multicolumn{5}{c}{\bf Table 1: Double Neutron Star Binaries$^{a,b,c}$} \\\hline
PSR & $P_b$ & $e$ & \multicolumn{1}{c}{measured PK parameter} & Reference \\\hline
J1518+4904 & 8.6 day & 0.25 & $\dot{\omega}$ & Nice, Sayer \& Taylor (1996) \nocite{nst96}  \\
{\bf B1534+12} & 10.1 hr & 0.27 & $\dot{\omega}, \gamma, \dot{P_b}, r, s$& see text  \\
J1811$-$1736 & 18.8 day & 0.83 & $\dot{\omega}$ & Lyne et al. (1999) \nocite{lcm+99} \\
{\bf B1913+16} & 7.8 hr & 0.62 & $\dot{\omega}, \gamma, \dot{P_b}$ & see text\\
B2127+11C & 8.0 hr & 0.68 & $\dot{\omega}$  & Prince et al. (1991)
\label{ta:psrs}
\end{tabular}
%\end{center}
{\footnotesize $^a$Sources suitable for tests of GR are indicated in bold.\\ 
$^b$PSR B1820$-$11 is not included as the nature of its companion is uncertain \\
\cite{pv91}.\\
$^c$PSR B2303+46, previously thought to have a neutron-star companion, has recently \\
been shown to have a white dwarf companion \protect\cite{vk99}.}
\end{table}

\section*{\psra}
\label{sec:1913}

The results of long-term timing observations of the relativistic binary
pulsar \psra\ are well-known; indeed they have been distinguished with the 1993
Nobel Prize in Physics awarded to the discoverers Joseph Taylor and
Russell Hulse.  Detailed descriptions and reviews of the results and
implications of those timing observations can be found in a variety of
references \cite{ht75a,thf+76,tw82,tay87,tw89,dt91,twdw92,dt92,tay92,tay93a}.  
Here we briefly summarize the status of those observations, and discuss the
recently reported evidence for geodetic precession in this system.

\subsection*{Status of Timing Observations of \psra}

As reported by Taylor (1993), \nocite{tay93a} timing observations of
the 59~ms \psra\ made at the 305~m radio telescope at Arecibo, Puerto
Rico through 1993 (the Arecibo telescope became inoperable not long
afterward in preparation for a major upgrade, which is nearly
complete) have resulted in the determination of three post-Keplerian
parameters: the rate of periastron advance $\dot{\omega} =
4^{\circ}.226621 \pm 0^{\circ}.000011$, the combined time dilation and
gravitational redshift $\gamma = 4.295 \pm 0.002$~ms, and the observed
orbital period derivative $\dot{P_b} = (-2.4225 \pm 0.0066) \times
10^{-12}$.  The first two of these parameters determine the component
masses to be $1.4411 \pm 0.0007$~$M_{\odot}$ and $1.3874 \pm
0.0007$~$M_{\odot}$.  The third post-Keplerian parameter, $\dot{P_b}$,
in principle allows for one test of GR (or other theory of gravity).

However, the observed value of $\dot{P_b}$ must first be corrected for the effect
of acceleration in the Galactic potential.  This correction follows from
the simple first-order Doppler effect, where $P_b^{obs} / P_b^{int} = 1 + v_R / c$,
where $P_b^{obs}$ and $P_b^{int}$ are the observed and intrinsic values, and
$v_R$ is the radial velocity of the pulsar relative to the solar system
barycentre.  A changing $v_R$ leads to a Galactic term
\begin{equation}
\left(\frac{\dot{P_b}}{P_b}\right) = \frac{a_R}{c} + \frac{v_T^2}{cd},
\label{eq:gal}
\end{equation}
where $a_R$ is the radial component of the acceleration, $v_T$ is the
transverse velocity, and $d$ is the distance to the pulsar.  The second
term in this equation is the familiar transverse Doppler or ``train-whistle''
effect.  The best estimate correction factor for \psra, given its only approximately
known location in the Galaxy, is $(-0.0124 \pm 0.0064) \times 10^{-12}$ \cite{dt91,tay92}.
With this correction applied to $\dot{P_b^{obs}}$, the comparison with
the GR prediction can be made; the result \cite{tay92} is that
\begin{equation}
\frac{\dot{P_b^{obs}}}{\dot{P_b}^{GR}} = 1.0032 \pm 0.0035.
\end{equation}
Note that the uncertainty in this expression is dominated by the uncertainty in
the Galactic acceleration term.
Since $a_R$ and $d$ are unlikely to be known with much greater
precision than
is currently available,
this particular test of GR will probably not improve much in the near future.

Additional tests of GR may still be possible with the \psra\ system if
the parameters $r$ and $s$ can be measured.  This may be possible
given the recent major upgrade to the Arecibo telescope, as higher timing
precision should now be available.

\subsection*{\psra\ and Geodetic Precession}

Relativistic geodetic precession, the gravitational analogue of Thomas
precession (the origin of fine structure in atomic spectra), is
predicted to result in a changing orientation of the pulsar spin axis.
As the pulsar precesses, our line of sight should intersect different
parts of the radio emission beam.  Thus, the average pulse profile
could vary significantly over time.  The first evidence for this
in the \psra\ system was presented by Weisberg, Romani \&
Taylor (1989) [but see also Cordes, Wasserman \& Blaskiewicz 1990].
\nocite{wrt89,cwb90}  They
reported a gradual, secular evolution in the ratio of the amplitudes of
the two pulse peaks (see Fig.~\ref{fig:std}).

\begin{figure}
\centerline{\epsfig{file=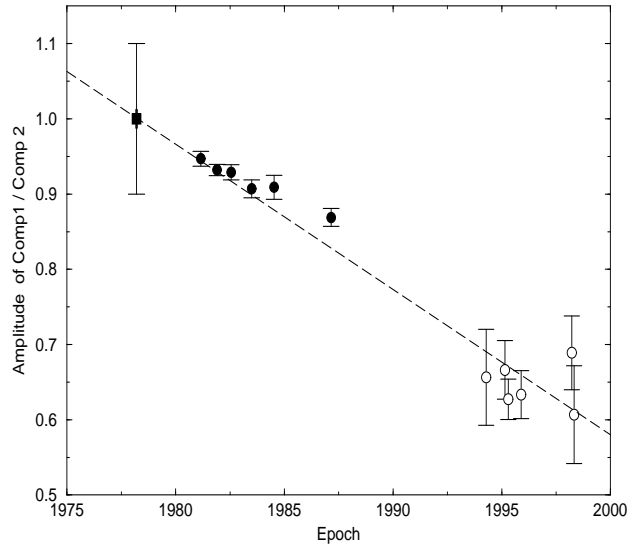,height=3.5in,width=3.5in,angle=270}}
\caption{The variation in the ratio of the amplitudes of the components
of the \protect\psra\ average radio pulse profile, after Kramer (1998).}
\label{fig:1913}
\end{figure}

Recently, Kramer (1998) \nocite{kra98} has clearly demonstrated that
this trend continues.  Figure~\ref{fig:1913} shows the ratio of the
amplitudes of the two pulse components as a function of time; the
variation is striking.  If the emission results from a cone of
radiation, then a secular change in the separation of the two peaks
ought to be observed as well; strong evidence for this is also now
seen \cite{kra98}.  Quantitative modeling of this variation depends on
the unknown beam morphology.  Under the assumption of a hollow,
circular emission beam, if GR is correct, Kramer shows that the
pulsar, sadly, will no longer grace the skies of our Earth after the
year 2025.  Happily however, it should reappear around the year 2220.
The exact dates of disappearance, together with the form of the
secular variation in average pulse morphology, will permit the first
direct observation and study of the morphology of a radio pulsar
emission beam.

\section*{\psrb}
\label{sec:1534}

The binary pulsar \psrb\ was discovered by Wolszczan (1991)
\nocite{wol91a} using the Arecibo telescope.  This 38~ms pulsar is in
a 10~hr eccentric orbit with a second neutron star (see Table~1).
\psrb\ offers the hope of additional and more precise tests of GR for
a number of reasons: first, the narrower pulse profile of \psrb\
(Fig.~\ref{fig:std}) means higher timing precision.  Second, the
orbital plane of this system is more inclined than that of \psra,
which facilitates the measurements of two additional relativistic
parameters $r$ and $s$.  Thus, in principle, five relativistic
post-Keplerian parameters are measurable with high precision for
\psrb, which allows two new additional tests of GR that have not been
done for \psra.  This is particularly important for testing
alternative theories of gravity, as it permits the separation of the
radiative and strong-field components of the theory.  This cannot be
accomplished in the simple $\dot{\omega}$-$\gamma$-$\dot{P_b}$ test,
as it mixes radiative and non-radiative effects
[see Damour \& Taylor 1992 for details].

\begin{figure}
\centerline{\epsfig{file=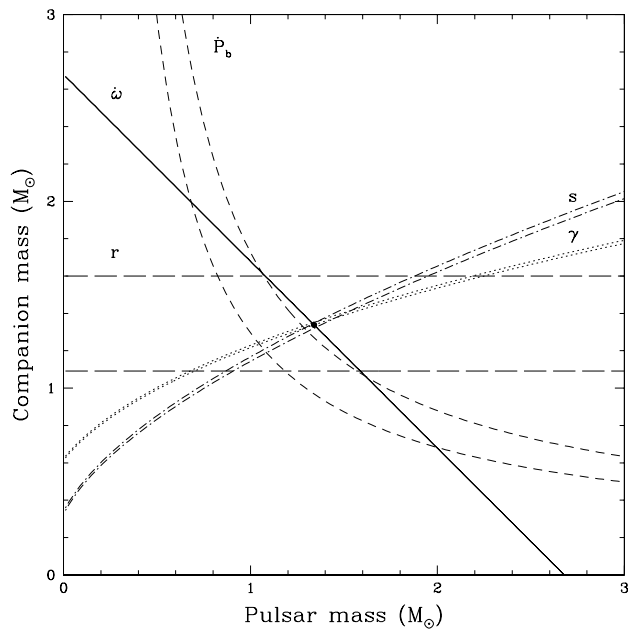,height=3.5in,width=3.5in}}
\caption{The constraints imposed by the observed values of the five
measured post-Keplerian parameters on the masses of the components
of \psrb, after Stairs et al. (1998).}
\label{fig:1534}
\end{figure}

Stairs et al. (1998) \nocite{sac+98} report on seven years of timing
observations of \psrb\ made at Arecibo, at the 43~m dish at Green
Bank, as well as at the 76~m Lovell radio telescope at Jodrell Bank.
As expected, they measure the five post-Keplerian relativistic
parameters $\dot{\omega}, \gamma, \dot{P_b}, r$ and $s$.  The results
are nicely summarized in Figure~\ref{fig:1534} (Fig.~4 in Stairs et
al. 1998), where the component masses are plotted on the axes.  As
each of the five post-Keplerian parameters has a different dependence
on the masses, each parameter defines a curve in this plane.  If GR
holds, then the five curves, as calculated in GR, should meet at a
single point.  As can be seen in Figure~\ref{fig:1534}, the curves for
$\dot{\omega}, \gamma$ and $s$ agree to better than 1\% (though that
for $r$ is not yet precise enough to be very constraining).
Surprisingly, their intersection implies that the pulsar and companion
have exactly equal masses within uncertainties, $1.339 \pm
0.003$~$M_{\odot}$.

As is clear in Figure~\ref{fig:1534}, the curve for $\dot{P_b}$ just
misses this intersection point.  Note, however, that the value of
$\dot{P_b}$ used to produce the curve in Figure~\ref{fig:1534}
included a correction for Galactic acceleration (Eq.~\ref{eq:gal})
that assumed a distance of 0.7~kpc to the pulsar, from its observed DM
and the best model for the free electron distribution \cite{tc93}.
The model is known to be only approximate, with uncertainties on
inferred distance for anyone source optimistically 25\%, and
realistically considerably larger.  Stairs et al.  therefore argue
that the discrepancy seen in Figure~\ref{fig:1534} can be removed by
simply invoking a larger distance to the pulsar, 1.1~kpc.  Put
differently, by assuming GR is correct, the distance to this
relativistic binary pulsar can be determined with greater precision
than is otherwise available \cite{bb96}.  This demonstrates that the
measurement of an improved $\dot{P_b}$ for \psrb\ is unlikely to offer
a useful test of GR unless the distance to the source can be
determined independently (for example, via a timing or interferometric
parallax measurement).  However, the expected improved determination
of the $r$ parameter, following the Arecibo upgrade, could yield a
useful test in addition to that from $\dot{\omega}$-$\gamma$-$s$.

The improved distance determination to \psrb\ made by Stairs et al. (1998) has
implications for estimates of the coalescence rate of double neutron star binaries.
A larger distance implies a more intrinsically luminous pulsar, which in turn
implies that there are fewer in the Galaxy, as otherwise more would be detected.
Stairs et al. suggest that the expected rate must be reduced relative to
previous estimates \cite{phi91,cl95,vl96} by factors of
2.5--20.  This rate is of considerable interest to the builders of gravitational
wave detectors like LIGO (see paper by P. Saulson, this volume).  Of course rates that
vary greatly depending on the estimated distance to a single object should be
regarded as crude estimates only.

\section*{Finding More Relativistic Binaries:\\ The Parkes Multibeam Survey}
\label{sec:1142}

A major survey of the Galactic Plane for radio pulsars is currently
underway.  This survey offers the hope of finding new examples of
relativistic binary pulsars suitable for studying GR effects.  The
observations are being done using the Parkes 64~m radio telescope in
Australia \cite{lcm+99}.  The survey is planned to cover the inner
Galactic Plane, in the Galactic longitude range $260^{\circ} < l <
50^{\circ}$ and Galactic latitude range $|b| < 5^{\circ}$.  The search
is being carried out at radio frequencies near 1400~MHz and has
roughly seven times the sensitivity of previous 1400~MHz surveys of
the Galactic Plane \cite{cl86,jlm+92}, owing mainly to the longer
integration time permitted by the use of the new multibeam receiver at
Parkes.  This new instrument consists of 13 independent,
non-overlapping receivers in the telescope focal plane.  This allows
the Galaxy to be surveyed to much greater depth than was previously
possible, without using a prohibitive amount of telescope time.  Each
beam pointing consists of a 35~min integration, with a total of
288~MHz bandwidth 1-bit sampled every 250~$\mu$s.  Some 35,000 beams
will be observed, and the data for each will be subject to a Fast
Fourier Transform of $2^{23}$ points.  The project is thus
computer resource intensive.  With approximately half of the survey
complete, 405 previously unknown radio pulsars have been discovered,
making this by far the most successful pulsar survey ever.

Among the first sources found in the survey is the very likely double neutron star
binary PSR~J1811$-$1736 (see Table~1) \cite{lcm+99}.  Although this system
is unlikely to be useful for tests of GR, its early discovery in the survey
suggests there are many more such systems to be found.  Indeed, not long
after the conference for which these proceedings are a record, 
a third relativistic binary pulsar suitable for tests of GR was discovered
among the new Parkes Multibeam sources.  Detailed observations of this
exciting source are just getting underway as this paper is being written.

\section*{Conclusions}

The now famous technique of timing relativistic binary pulsars has
yielded confirmation that GR is the correct theory of gravity at
better than the 1\% level.  Future additional tests of GR, using the
only two known sources well-suited to such tests, \psrb\ and \psra,
are possible, from improved measurements of the Shapiro delay $r$ and
$s$ parameters.  The precision in the
$\dot{\omega}$-$\gamma$-$\dot{P_b}$ test is limited by the uncertainty
in our estimates for the Galactic acceleration of these objects.
However, under the now justified assumption that GR is correct,
observations of relativistic binary pulsars can yield unique
astrophysical measurements that have never before been possible,
including precise determination of neutron star masses, distances to
these sources, LIGO source rates, and morphological studies of the
pulsar radio emission beam.  The ongoing Parkes Multibeam survey of
the Galactic Plane promises (and indeed has already begun) to discover
new examples of these fascinating objects.

\bigskip

VMK is an Alfred P. Sloan Research Fellow.  She thanks Michael Kramer
and Ingrid Stairs for sharing their figures, and the organizers of the
8th Canadian Conference on General Relativity and Relativistic
Astrophysics for their hospitality and patience.

%\bibliographystyle{/nfs/janeway/h1/vicky/doc/tex/apj1c}
%\bibliography{/nfs/janeway/h1/vicky/doc/tex/journals1,/nfs/janeway/h1/vicky/doc/tex/modrefs,/nfs/janeway/h1/vicky/doc/tex/psrrefs}

\end{document}